\documentclass{nature2}

\usepackage{astjnlabbrev-nature}

\usepackage{epsfig}
\usepackage[usenames,dvipsnames]{xcolor}
\usepackage{graphicx}
\usepackage[caption=false]{subfig}
\usepackage{longtable}
\usepackage{bm}
\usepackage{hyperref}

\hypersetup{
    colorlinks=true,
    linkcolor=blue,
    citecolor=blue,      
    urlcolor=blue,
    }

\usepackage{enumitem}
\setlist{nolistsep} 
\setitemize[0]{leftmargin=*}
\usepackage{multirow}
\usepackage{colortbl}
\usepackage{tabulary}
\usepackage{amsmath}
\usepackage{bm}
\usepackage{amssymb}
\usepackage{tikz,lipsum,lmodern}

\usepackage{hanging}

\newcommand{\ie}{i.e.,}

\newcommand{\spitzer}{{\it Spitzer}}

\newcommand{\wise}{{\it WISE}}
\newcommand{\galex}{{\it GALEX}}

\newcommand{\herschel}{{\it Herschel}}
\newcommand{\iras}{{\it IRAS}}
\newcommand{\nastro}{Nat. Astron.} 

\newcounter{daggerfootnote}
\newcommand*{\daggerfootnote}[1]{%
	\setcounter{daggerfootnote}{\value{footnote}}%
	\renewcommand*{\thefootnote}{\fnsymbol{footnote}}%
	\footnote[2]{#1}%
	\setcounter{footnote}{\value{daggerfootnote}}%
	\renewcommand*{\thefootnote}{\arabic{footnote}}%
}

\colorlet{juan}{JungleGreen!70!RoyalBlue}

\usepackage{longtable}
\usepackage[labelsep=endash]{caption}

\title{Merger-driven infall of metal-poor gas in luminous infrared galaxies: a deep dive beneath the mass-metallicity relation}

\author{Borja P\'{e}rez-D\'{\i}az$^{1}$, Enrique P\'{e}rez-Montero$^{1}$, Juan A. Fern\'{a}ndez-Ontiveros$^{2}$,
  Jos\'{e} M. V\'{\i}lchez$^{1}$, Ricardo Amor\'{\i}n$^{3, 4}$}

\makeatletter
\let\saved@includegraphics\includegraphics
\AtBeginDocument{\let\includegraphics\saved@includegraphics}
\makeatother

\begin{document}
\maketitle
\begin{affiliations}
\item {\small Instituto de Astrof\'{\i}sica de Andaluc\'{\i}a (IAA-CSIC), Glorieta de la Astronom\'{\i}a s/n, 18008 Granada, Spain}
\item {\small Centro de Estudios de F\'{\i}sica del Cosmos de Arag\'{o}n (CEFCA), Unidad Asociada al CSIC, Plaza San Juan 1, E--44001 Teruel, Spain}
\item {\small Instituto de Investigaci\'{o}n Multidisciplinar en Ciencia y Tecnolog\'{\i}a, Universidad de La Serena, Raul Bitr\'{a}n 1305, La Serena, Chile}
\item {\small Departamento de Astronom\'{\i}a, Universidad de La Serena, Av. Juan Cisternas 1200 Norte, La Serena, Chile}

\end{affiliations}

\begin{abstract}
The build up of heavy elements and the stellar mass assembly are fundamental processes in the formation and evolution of galaxies. Although they have been extensively studied through observations and simulations, the key elements that govern these processes, such as gas accretion and outflows, are not fully understood\cite{Maiolino_Mannucci_2019}. This is especially true for luminous and massive galaxies, which usually suffer strong feedback in the form of massive outflows\cite{Arribas_2014, Cazzoli_2016, Cicone_2014, Pereira-Santaella_2018}, and large-scale gas accretion triggered by galaxy interactions\cite{Joseph_Wright_1985, Howell_2010, Kennicutt_Evans_2012}.
For a sample of 77 luminous infrared (IR) galaxies, we derive chemical abundances using new diagnostics based on nebular IR lines, which peer through the dusty medium of these objects and allow us to include the obscured metals in our abundance determinations. In contrast to optical-based studies, our analysis reveals that most luminous IR galaxies remain close to the mass-metallicity relation. Nevertheless, four galaxies with extreme star-formation rates ($> 60$M\ensuremath{_{\odot }}yr\ensuremath{^{-1}}) in their late merger stages show heavily depressed metallicities of 12+log(O/H)\,$\sim 7.7$--$8.1$ along with solar-like N/O ratios, indicative of gas mixing processes affecting their chemical composition. This evidence suggests the action of a massive infall of metal-poor gas in a short phase during the late merger stages, eventually followed by a rapid enrichment. These results challenge the classical gas equilibrium scenario usually applied to main-sequence galaxies, suggesting that the chemical enrichment and stellar-mass growth in luminous IR galaxies are regulated by different processes.
\end{abstract}

The chemical composition of the gas-phase interstellar medium (ISM) is a witness of the evolution of galaxies, as heavy elements (or metals) are produced in stars by stellar nucleosynthesis and eventually ejected into the ISM at the end of their lives\cite{Nomoto_2013}. As stellar mass assembly also traces the evolution and formation of galaxies, a natural connection between them arises, materialized in the so-called Mass-Metallicity Relation (MZR)\cite{Lequeux_1979, Tremonti_2004, Andrews_Martini_2013, Perez-Montero_2016} as well as in the Fundamental Metallicity Relation (FZR)\cite{Mannucci_2010, Curti_2020}, which are regulated by a wide variety of processes: galaxy environment\cite{Peng_Maiolino_2014}, secular evolution\cite{Somerville_Dave_2015}, feedback from star-formation\cite{Blanc_2019} and Active Galactic Nuclei (AGN)\cite{Thomas_2019}, and stellar age\cite{Duarte-Puertas_2022}.

These scaling relations are explained by means of self-regulated equilibrium between gas accretion and star formation\cite{Maiolino_Mannucci_2019}. However, half of the stellar mass in galaxies that we observe nowadays was formed in a relatively short period of time (\ensuremath{\sim } 3.5 Gyr) during the cosmic noon\cite{Forster_Wuyts_2020}, i.e., in an extreme scenario where these equilibrium conditions may not apply. Luminous and Ultra-Luminous Infrared Galaxies (LIRGs and ULIRGs, respectively) are key to understanding the complex picture of galaxy evolution, since they allow us to explore nearby galaxies with extreme star formation rates, similar to those found at the cosmic noon\cite{Armus_2009}. LIRGs have infrared luminosities L\ensuremath{_{\mathrm{IR}}>} 10\ensuremath{^{11}}L\ensuremath{_\odot} by definition, dominate the star-formation activity at z\ensuremath{\sim}1 and contribute significantly to the cosmic infrared background\cite{Stierwalt_2013}. In addition, ULIRGs with L\ensuremath{_{\mathrm{IR}}>} 10\ensuremath{^{12}}L\ensuremath{_\odot} are commonly found in merger systems at z\ensuremath{\gtrsim}2 and may represent an evolutionary stage in the formation of AGN-hosting galaxies\cite{Sanders_1988, Evans_2005}. While the scaling relation between the bulge and central supermassive black holes (SMBHs) in quiescent galaxies suggest a link between accretion rates onto SMBH and the intensity of the episodes of nuclear star formation, \cite{Gebhardt_2000, Magnelli_2009}, the study of (U)LIRGs and their extreme conditions are critical to shed light on the co-evolution of SMBHs and their host galaxies.

Despite the increasing number of studies on the gas-phase ISM metal content in galaxies, the chemical characterization of ULIRGs lacks consensus. On the observational side, optical estimations of the oxygen abundance\cite{Caputi_2008, Rupke_2008} favour a low-metal content gas scenario for these galaxies. However, this approach
presents serious challenges: 1) optical emission lines are significantly affected by dust attenuation, which is particularly important in (U)LIRGs\cite{Franceschini_2003}, and therefore miss the heavy elements located in dust-embedded regions within galaxies; 2) due to the hard ionising radiation field associated with strong starbursts, the cooling process of many metals (such as sulfur, neon or oxygen) may be dominated by highly-ionized species (S$^{3+}$, Ne$^{2+}$, and even O$^{3+}$ in some extreme scenarios), whose emission lines are detected in the IR range; and, 3) due to the temperature dependence of optical emission lines, cold regions might remain unobserved. To avoid these problems, studies of the metallicity of ULIRGs from IR emission lines are needed, but such studies are rather scarce and show discrepancies. A study of local ULIRGs using far-IR emission lines was performed on a sample of 20 local galaxies\cite{Pereira-Santaella_2017}, concluding that 12+log(O/H) ranges from 8.5 to 8.9 [0.65Z\ensuremath{_{\odot}}, 1.65\ensuremath{Z_{\odot}}], with many of them found $0.3\, \rm{dex}$ lower than expected by their position in the MZR. More recently, another study of 5 ULIRGs using again far-IR emission lines showed that, contrary to the previous study, these ULIRGs followed the MZR\cite{Chartab_2022} derived for SFGs\cite{Tremonti_2004} in the Sloan Digital Sky Survey (SDSS). However, as these studies were performed using ratios of far-IR emission lines (including [O{\sc iii}]\ensuremath{\lambda}52\ensuremath{\mu}m, [N{\sc iii}]\ensuremath{\lambda}57\ensuremath{\mu}m, [O{\sc iii}]\ensuremath{\lambda}88\ensuremath{\mu}m) that are better tracers of the nitrogen-to-oxygen abundance ratio log(N/O)\cite{Peng_2021, FO21}, the estimation of 12+log(O/H) relies on the local calibration of the O/H--N/O relation\cite{PMContini_2009, PM14}, that might not apply to the case of (U)LIRGs since independent estimations of both quantities lead to subsolar abundances\cite{FO21}. On the theoretical side, due to the high dust content\cite{Herrero-Illana_2019} of ULIRGs, chemical evolution models predict high metal content in the ISM of these galaxies\cite{Calura_2006}. On the other hand, simulations of ULIRGs, as interacting systems, have shown that the chemical content in the gas-phase ISM, mainly traced by the oxygen abundance (12+log(O/H)), experiences a drop during the merger process\cite{Montuori_2010, Rupke_2010}. Overall, the lack of a systematic and consistent study of the chemical content in the gas-phase ISM of ULIRGs does not help in solving this problem since an independent determination of the O/H and N/O abundances based on IR emission lines is required to peer through the dusty medium of these galaxies. 

\begin{figure}[h!]
	\centering
	\includegraphics[scale=0.65]{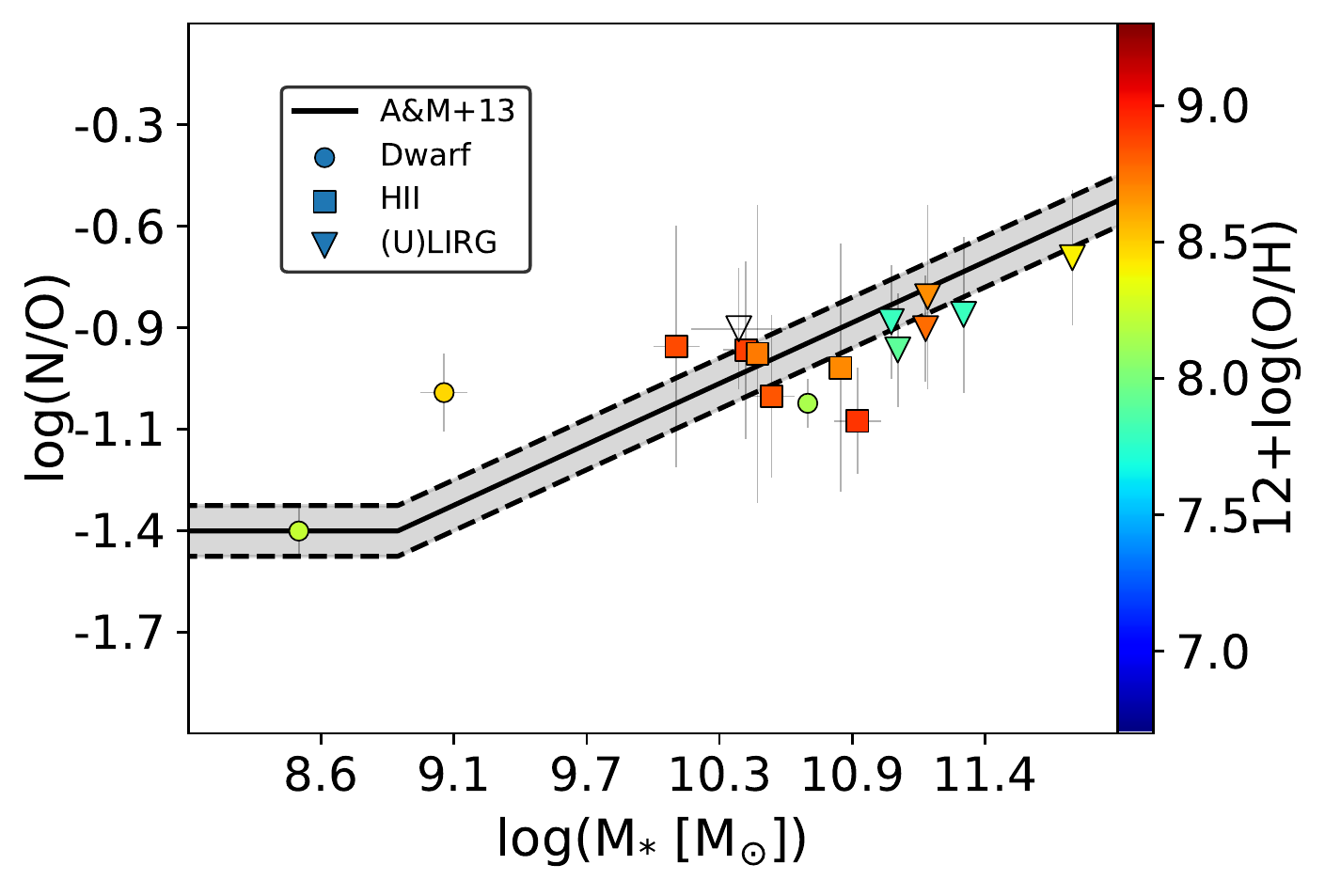} 
	\caption{Relation between stellar mass and nitrogen-to-oxygen ratio for our sample of galaxies. The local relation found for SFGs\cite{Andrews_Martini_2013} is shown.}
	\label{fig:no}
\end{figure}

In this study, we determine the chemical abundances of the ISM in a sample of 77 (U)LIRGs dominated by star-formation activity, and we use a sample of 55 HII extragalactic regions and star-forming dwarf galaxies as a control sample. Thus, a total sample of 132 galaxies with IR observations is analysed.  We use spectroscopic observations of nebular emission lines in the near-, mid- and far-IR ranges acquired with \textit{Spitzer}/IRS, \textit{Herschel}/PACS, and \textit{Akari} for measurements of the Br$\alpha$ emission. To estimate 12+log(O/H), log(N/O) and the ionization parameter log(U) we use {\sc HII-CHI-Mistry-IR}\cite{FO21, PD22}, a code that employs a bayesian-like comparison between the results predicted by large grids of photoionization models with specific observed emission-line ratios. This code allows us to independently estimate these three quantities without assuming any underlying relation between any of them. However if the set of emission lines used as input is reduced (e.g. when no previous estimation of N/O can be provided), the code assumes certain relations that can be changed by the user to find an estimation of the chemical content. For our sample, we use IR emission lines from H{\sc I}\ensuremath{\lambda}4\ensuremath{\mu }m to [O{\sc iii}]\ensuremath{\lambda}88\ensuremath{\mu}m as input for the code (see Methods for more details). Overall, we find that most galaxies in our sample have a nearly solar metallicity 12+log(O/H)\ensuremath{\sim}8.6 (0.85Z\ensuremath{_{\odot}}), which is consistent with previous studies from IR\cite{Pereira-Santaella_2017} emission lines. We find a median N/O value for our sample of \ensuremath{\sim}-1.0, that is consistent with the expected values for their stellar masses (see Figure \ref{fig:no}).

\begin{figure}[ht!!]
	\centering
	\includegraphics[scale=0.55]{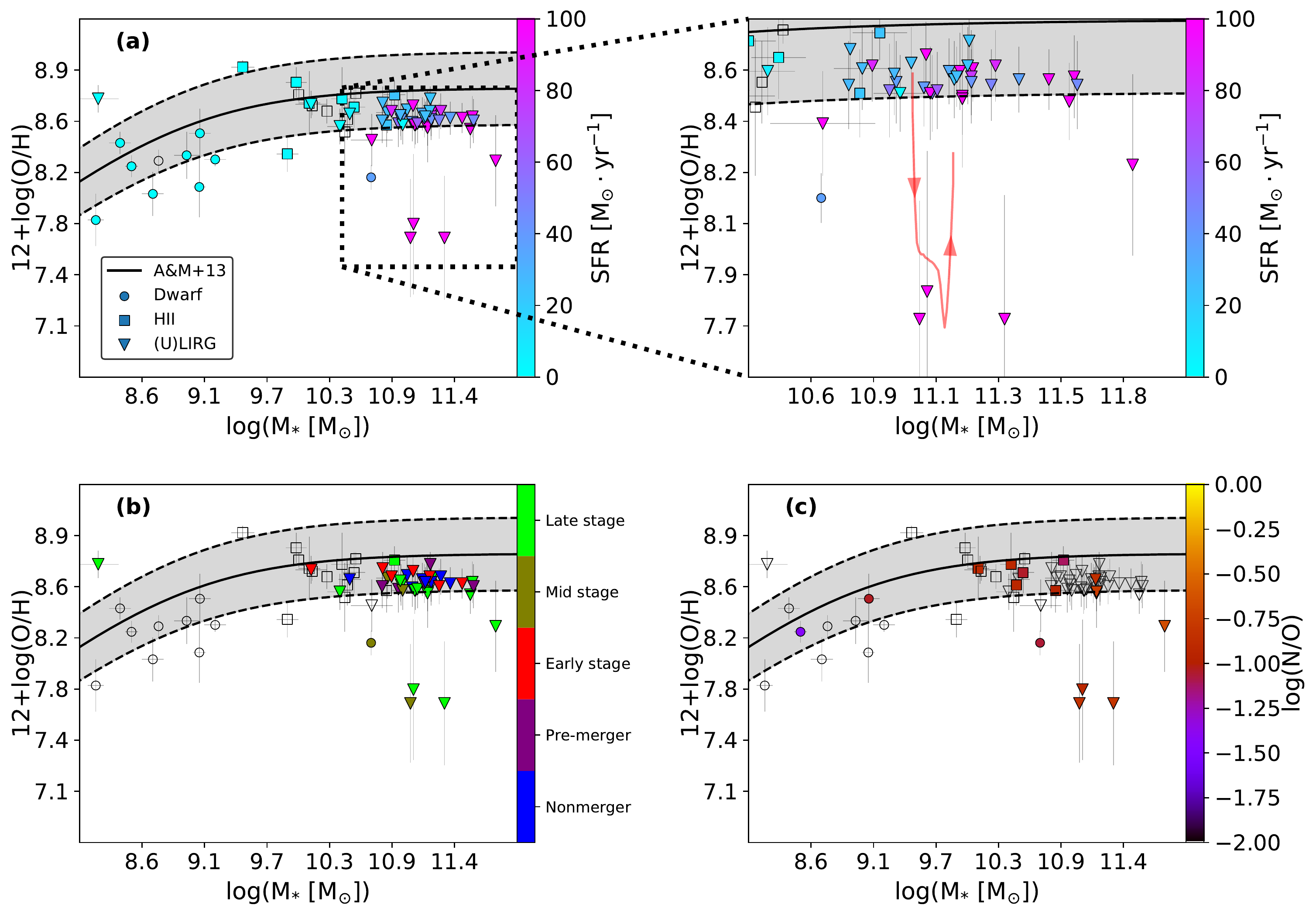} 
	\caption{Mass-metallicity relation for our sample of SFGs. The colorbar shows different properties for their host galaxies: {\it a)}, the star-formation rate (SFR) as retrieved from the literature; {\it b)}, the merger stage as estimated from IRAC 3.6\ensuremath{\mu}m images; {\it c)} the N/O chemical ratio. For all plots, blank points are associated with galaxies with no estimations of the colored properties. For plot {\it a)} we show on the right a zoom with tracks (red line) of an inflow model within the standard evolution of a galaxy\cite{Koppen_Edmunds_1999} (see Methods for more details on this model).}
	\label{fig:MZR}
\end{figure}

We present in Figure \ref{fig:MZR} the MZR for 77 galaxies, including 41 (U)LIRGs, using the oxygen abundances estimated in this work. We compare our results with the MZR previously reported for SFGs\cite{Andrews_Martini_2013} (A\&M13 track).
We find that the large majority of our sample follows the same trend as the rest of SFGs. However, the most massive ULIRGs (\ensuremath{\log(M_{*} [M_{\odot}]) > 10.5}) tend to lie towards the low side of the MZR, indicating that they have slightly lower metallicity as compared to the expected value for their stellar mass, confirming what has been also reported  in previous studies \cite{Pereira-Santaella_2017}, although the deviations that we find  are smaller than the values found by other studies\cite{Rupke_2008}. Interestingly, four galaxies namely Haro11, IRAS12112+0305, IRAS20551-4250 and IRAS23128-5919 strongly deviate from the MZR showing metallicities lower by a factor of \ensuremath{\gtrsim}2 than expected for these {\it deep-diving} (U)LIRGs stellar masses according to the MZR. While Haro11 is not a ULIRG galaxy like the others, this star-forming galaxy is reported to be a well known merger dominated galaxy that shared many of properties\cite{Ostlin_2015, Ostlin_2021} observed in ULIRGs. Notably, IRAS12112+0305 was previously reported to have a super-solar abundance (\ensuremath{\sim}2Z\ensuremath{_{\odot}})\cite{Chartab_2022}, while IRAS20551-4250 and IRAS23128-5919 where reported to have abundances close to the values expected from the MZR ar their masses\cite{Herrera-Camus_2018}. These high estimations of O/H are driven by the nearly solar N/O values obtained  here. However, our O/H estimation -- independent of N/O-- is significantly lower, thus suggesting a more complex mixing process of the gas affecting their chemical composition. These deviations from the local relation between N/O--O/H found in SFGs have been also reported for other types of galaxies such as {\it Green Pea} Galaxies\cite{Amorin_2010}.

\begin{figure}[ht!]
	\centering
	\includegraphics[scale=0.60]{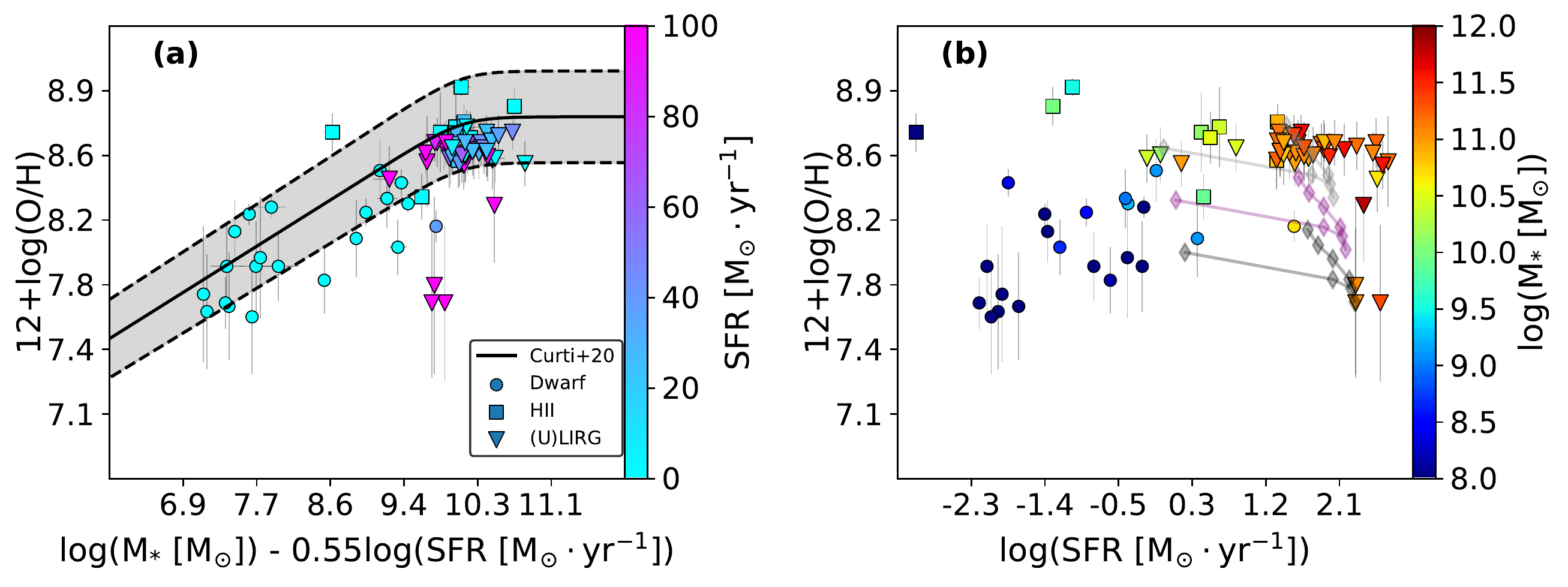} 
	\caption{Relations between SFR, stellar mass and metallicity. {\it a)} shows the FMR relation for log(M$_{*}$ [M$_{\odot}$]) - $\alpha$log(SFR [M$_{\odot}\cdot$yr$^{-1}$]) assuming $\alpha =  0.55$\cite{Curti_2020}. {\it b)} shows the relation between O/H and SFR for our sample of galaxies, while semi-transparent lines represent tracks obtained in simulations of mergers of galaxies with the same stellar mass\cite{Montuori_2010}, considering different initial conditions.}
	\label{fig:SFR_FMR}
\end{figure}

We show in Figure \ref{fig:MZR} that these {\it deep-diving} (U)LIRGs have the highest SFR values in our sample (\ensuremath{>100} M\ensuremath{_{\odot}}yr\ensuremath{^{-1}}) and they are in the final stages of their merges. Additionally, we find that the N/O ratios for these four galaxies are similar to the solar ratio, which suggests that they have reached a relatively mature stage of their chemical enrichment history\cite{Amorin_2010, Perez-Montero_2013}. These results imply that during the merging process of (U)LIRGs, metal-poor gas is accreted towards the central region of the galaxies, which dilutes the metal abundances relative to hydrogen, fuels star formation, but keeps the N/O ratio unaffected because it is independent of the effects of a massive infall of hydrogen. Additionally, intermediate-massive stars of these galaxies do not have enough time to boost the nitrogen production from CNO cycles\cite{Vincenzo_Kobayashi_2018}. As also shown in Figure \ref{fig:SFR_FMR} {\it a)}, these {\it deep-diving} (U)LIRGs also deviate from the FZR relation, implying that these (U)LIRGs are mainly out of the self-regulated equilibrium between gas accretion and star formation.

To strengthen the case for massive infall of gas in these {\it deep-diving} (U)LIRGs, we discuss two models of inflow that reproduce the trends of these inflows. In Figure \ref{fig:MZR} {\it a)}, we present a basic infall model\cite{Koppen_Edmunds_1999}, in which the analyzed galaxy (or common envelope of two galaxies) experiences a drop of metallicity over a period of time between 100 to 1000 Myr. These timescales further support our scenario and explain why only four galaxies ({\it deep-diving} (U)LIRGs) in our sample were captured in the process, as these timescales are really short for a {\ensuremath \sim} 12 Gyr old-like galaxy. Once the merger process is completed and stars begin to pollute the ISM again, the resulting galaxy increases its metallicity. According to this model, an additional contribution from outflows is necessary for the merger to recover the metallicity expected from the MZR for a galaxy of that stellar mass. This is in line with the metal-loaded outflows often found in ULIRGs\cite{Arribas_2014, Cazzoli_2016, Cicone_2014, Pereira-Santaella_2018}. In Figure \ref{fig:SFR_FMR} {\it b)} we present tracks that correspond to different stages in a merger process between two galaxies of similar masses, as inferred from simulations\cite{Montuori_2010} in the SFR-O/H relation which is known to play an important role in the MZR\cite{Mannucci_2010, Maiolino_Mannucci_2019, Curti_2020, Duarte-Puertas_2022}. While these trends are dependent on the initial conditions of the involved galaxies, they all point towards a drop in metallicity which corresponds to a period of time of 600-850 Myr during the merger process\cite{Montuori_2010}, consistentlyy with the times derived for our infall model.

In summary, our study reveals that (U)LIRGs exhibit {\it loops} in the MZR due to the rapid infall of massive clouds of metal-poor gas that promote violent star formation in these galaxies. This scenario explains several observed properties of (U)LIRGs, including the discrepancy among different chemical models in reproducing the dust content of these galaxies, the required fuel to sustain their high star formation rates, and the anomalous low metal content of mid- and late-merger stage galaxies 
despite their high N/O ratios indicative of an advanced chemical stage. Both simulations and simple inflow models can reproduce this scenario, and once the infall phase is complete, the role of outflows is crucial for the galaxy to regain its initial metal content. Under these conditions, the build up of heavy elements and the stellar mass growth are not self-regulated by the same gas equilibrium prescription applied to main-sequence galaxies. Overall, due to the short timescales during which the infall phase can be captured, only a few galaxies in our sample demonstrate clear evidence of this scenario.

\newpage

\begin{addendum} 

\item[Acknowledgements]
BPD, EPM and JVM acknowledge financial support from the grant CEX2021-001131-S funded by MCIN/AEI/ 10.13039/501100011033. BPD, EPM and JVM also acknowledge support from the Spanish MINECO grants AYA2016-76682C3-1-P, AYA2016-79724-C4 and PID2019-106027GB-C41. JAFO acknowledges the financial support from the Spanish Ministry of Science and Innovation and the European Union -- NextGenerationEU through the Recovery and Resilience Facility project ICTS-MRR-2021-03-CEFCA, and the grant PDG2021-124918-NB-C44. RA acknowledges support from ANID Fondecyt Regular 1202007. EPM acknowledges the assistance from his guide dog Rocko without whose daily help this work would have been much more difficult.

\item[Author Contributions] {BPD, EPM and JAFO authored the draft version of this paper. BPD and JAFO compiled the sample of galaxies from the literature as well as all the ancillary data required for the analysis. EPM and BPD developed the code used to estimate chemical abundances. JMV and RA supervised the models and contributed to interpreting the results and contributed to the improvement of this manuscript.}

\item[Correspondence] 
{Correspondence for materials should
be addressed to B. P.-D. (bperez@iaa.es), E. P.-M. (epm@iaa.es) and J. A. F.-O. (j.a.fernandez.ontiveros@gmail.com)}

\item[Competing interests statement] The authors declare no competing interests.

\end{addendum}

\clearpage

\begin{center}
{\bf \large Methods}
\end{center}
\vspace{-0.6cm}

Here we provide more detailed information in the selection of the sample (\S~\ref{sec:sample}) of Ultra-Luminous Infrared Galaxies (ULIRGs) and Star-forming Galaxies (SFGs) with IR spectroscopic data. We also described the methodology employed to estimate the chemical content of the gas-phase interstellar medium (ISM) in our sample (\S~\ref{sec:chemical}). In addition, we give some details about the retrieval of other properties derived in the sample, such as the stellar mass (M\ensuremath{_{*}}), the star-formation rate (SFR), or the merger stage (\S~\ref{sec:ancillary}). Finally, we present in detail some evolutionary models that reproduce the behaviour of our sample of galaxies (\S~\ref{sec:models}). Throughout this study, we assume as reference system the solar chemical abundances\cite{Lodders_2021}.

\section{The sample of Ultra-luminous Infrared Galaxies and Star-forming Galaxies}
\label{sec:sample}
\subsection{Selection}
We build our sample of galaxies with IR spectroscopic data from two different catalogs. The first one was retrieved from the IDEOS\cite{HernanCaballero_2016, Spoon_2022} IR database, which includes the measurements of 77 fitted mid-IR observables in the range 5.4-36 $\mu$m for all galaxies observed with \spitzer\ (a total of 3335 galaxies\cite{Spoon_2022}). From this initial sample, we omit duplicate observations of one single object and entries corresponding to galaxies presented in the first catalog, where the 
higher resolution spectroscopic observation was taken. After verifying that star formation dominates the ionisation budget in our final sample of galaxies (see next subsection), we perform a cross-match with an {\it AKARI} sample of ULIRGs\cite{Inami_2018} to provide measurements of at least one hydrogen recombination line, particularly H{\sc i} (5-4). In all, a total sample of 66 ULIRGs was retrieved.

The second catalog\cite{FO21} consists of a sample of dwarf galaxies, SFGs, and ULIRGs that present mid- to far-IR range spectroscopic observations from the InfraRed Spectrograph (IRS\cite{Houck_2004}) on board \spitzer\ and from the Photodetector Array Camera and Spectrometer (PACS\cite{Poglitsch_2010}) on board \herschel\cite{FO21, FO16}. We compiled measurements for a sample of 66 galaxies showing detections of [O{\sc iii}]\ensuremath{\lambda}52\ensuremath{\mu}m and [N{\sc iii}]\ensuremath{\lambda}57\ensuremath{\mu}m far-IR lines\cite{Peng_2021} from the IFU spectroscopy instrument FIFI-LS\cite{Fischer_2018} on board the SOFIA airborne observatory. Additionally, we obtained measurements of hydrogen recombination lines H{\sc i} (7-6) and (5-4) from the \spitzer\ calibrated spectra available in the CASSIS\cite{Lebouteiller_2015} database and from observations via AKARI/IRC\cite{Imanishi_2010}, respectively.

\subsection{Classification}

For the first sample of galaxies, the ratio between [Ne{\sc v}]\ensuremath{\lambda}14\ensuremath{\mu}m and [Ne{\sc ii}]\ensuremath{\lambda}12\ensuremath{\mu}m ensures\cite{Pereira-Santaella_2017, FO21} an AGN contamination lower than 10\%. For the second sample, we use two criteria\cite{Sturm_2002, Tommasin_2010} both based on the relative emission of high ionic species ([Ne{\sc v}]\ensuremath{\lambda}14\ensuremath{\mu}m and [O{\sc iv}]\ensuremath{\lambda}26\ensuremath{\mu}m) to low ionic species ([Ne{\sc ii}]\ensuremath{\lambda}12\ensuremath{\mu}m) and on the strength of the  polycyclic aromatic
hydrocarbon (PAH) at 11.25\ensuremath{\mu}m, since AGN activity is expected to enhance higher ionic species\cite{PD21, PD22} and the equivalent width (EW) of PAHs has been proposed\cite{Puget_Leger_1989} as a tracer of the star-formation activity.

\begin{figure}[h]
    \centering
    \includegraphics[scale=0.65]{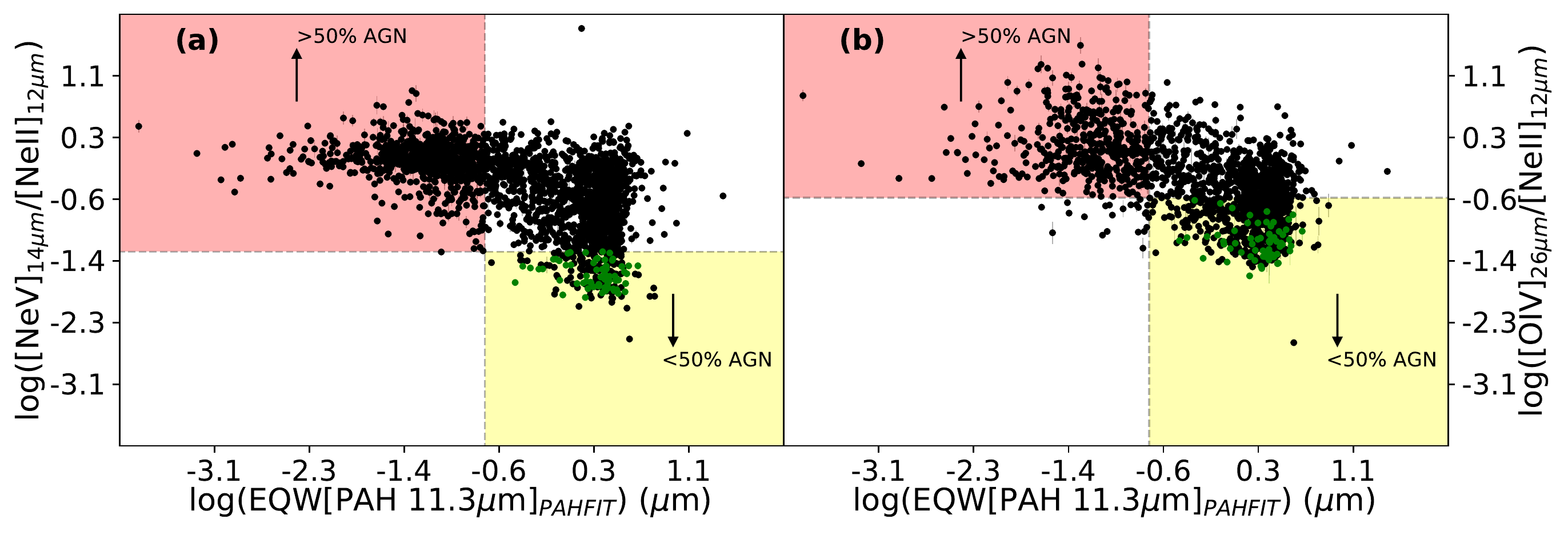} 
    \caption{Diagnostic diagrams for the AGN contamination\cite{Sturm_2002, Tommasin_2010} based on IR spectral features. The whole sample of galaxies from IDEOS is presented in these plots, while the sample of (U)LIRGs retrieved is highlighted in green. {\it a)} Diagnostic diagram using [Ne{\sc v}]\ensuremath{\lambda}14\ensuremath{\mu}m as a high-ionized specie. {\it b)} Diagnostic diagram using [O{\sc iv}]\ensuremath{\lambda}26\ensuremath{\mu}m instead.}
    \label{fig:SFG_diagnostic}
\end{figure}

In Figure \ref{fig:SFG_diagnostic} we present both criteria for the sample of galaxies retrieved from IDEOS. To ensure the robustness in the classification and avoid AGN contamination in first order, we only select galaxies which belong to the SF-dominated region (yellow shaded region) in both diagrams. Moreover, when compared our classification to that presented in IDEOS\cite{Spoon_2022}, we found that the whole selected sample (66 galaxies) show silicate strength and an equivalent width for the PAH at 6.2\ensuremath{\mu}m also compatible with star-forming dominated activity.

\section{Chemical abundance estimations}
\label{sec:chemical}
The estimation of chemical abundances, the oxygen content (12+log(O/H)) and the nitrogen-to-oxygen ratio (log(N/O)), as well as the ionization parameter (log(U)) was performed using {\sc HII-CHI-Mistry-IR}\cite{FO21,PD22} (or {\sc HCm-IR}). In brief, {\sc HCm-IR}\daggerfootnote{All versions of the {\sc HII-CHI-Mistry} code are publicly available at: \url{http://www.iaa.csic.es/~epm/HII-CHI-mistry.html}.} is a python code that performs a Bayesian-like calculation of the above mentioned quantities by comparing a set of emission lines with the predictions calculated from a grid of photoionization models\cite{PM14}. By default, the code provides four different grids of models, varying the Spectral Energy Distribution (SED) used as ionization source. For our study, since our sample is dominated by star-formation, we use the {\sc POPSTAR} grid, which was computed assuming a SED of simple stellar population models from {\sc POPSTAR}\cite{Molla_2009}, assuming a constant density of 100 cm$^{-3}$, a filling factor of 0.1 and plane-parallel geometry. Previous studies of the code have shown that these assumptions do not introduced large uncertainties in the derived abundances, although some differences can be found in the ionization parameter\cite{PM14, PM17, PM19, PD21, PD22, PM23}. The grid covers a range in 12+log(O/H) from 6.9 to 9.1 in bins of 0.1 dex, in log(N/O) from -2.0 to 0.0 in bins of 0.125 dex and in log(U) from -4.0 to 1.5 in bins of 0.25 dex.

The lines accepted by the code in the input include
 H{\sc i}\ensuremath{\lambda}4.05\ensuremath{\mu}m, [Ar{\sc ii}]\ensuremath{\lambda}6.99\ensuremath{\mu}m, H{\sc i}\ensuremath{\lambda}7.46\ensuremath{\mu}m, [Ar{\sc iii}]\ensuremath{\lambda}8.99\ensuremath{\mu}m, H{\sc i}\ensuremath{\lambda}12.4\ensuremath{\mu}m, [Ne{\sc ii}]\ensuremath{\lambda}12.8\ensuremath{\mu}m, [Ne{\sc v}]\ensuremath{\lambda}14.3\ensuremath{\mu}m, [Ne{\sc iii}]\ensuremath{\lambda}15.6\ensuremath{\mu}m, [Ne{\sc v}]\ensuremath{\lambda}24\ensuremath{\mu}m, [O{\sc iv}]\ensuremath{\lambda}26\ensuremath{\mu}m, [O{\sc iii}]\ensuremath{\lambda}52\ensuremath{\mu}m, [N{\sc iii}]\ensuremath{\lambda}57\ensuremath{\mu}m, [O{\sc iii}]\ensuremath{\lambda}88\ensuremath{\mu}m.
As compared to previous versions of the code employed for other samples of SFGs\cite{FO21} or AGNs\cite{PD22}, we introduce two modifications (version 3.1). The first modification is that the input now accepts argon emission lines, which behave in the same way as neon emission lines. Secondly, the code does no longer use sulfur emission lines to derive the oxygen abundance ratio as they are used instead to estimate the sulfur content independently from the oxygen estimation. Some authors\cite{PM06, DiazZamora_2022} have pointed that sulfur-to-oxygen ratios might deviate from the solar ratio in the low- and high-metal regime, implying that the use of sulfur emission lines to estimate the oxygen content might introduce a bias. The code finds solutions for  12+log(O/H), log(N/O) and log(U) as the average of the \ensuremath{\chi^{2}}-weighted distribution of all models, being \ensuremath{\chi^{2}} the quadratic differences between the observed and the predicted emission-line ratios sensitive to the above chemical ratios built upon the emission lines cited above. The code performs an independent estimation of log(N/O) and 12+log(O/H), although if N/O cannot be constrained due to the lack of key emission lines, a relation between O/H and N/O is assumed by the code. By default, this relation is the one obtained for star-forming regions using chemical abundances based on  optical emission-lines\cite{PMContini_2009, PM14}. In this work, as N/O can be estimated independently from far-IR emission lines, this relation is not used.

\section{Ancillary data}
\label{sec:ancillary}
To perform our study, we have retrieved almost all properties that are key in understanding the chemical evolution of the galaxies, \ie the stellar mass, the star-formation rate or the merger state. In this section, we detail how we obtained these values from the literature and the methodologies followed to estimate them.
\subsection{Stellar mass and star formation rates}
For 30 galaxies, we retrieved their stellar masses and star formation rates from the Great Observatories All-sky LIRG Survey (GOALS\cite{Armus_2009, Howell_2010}). Particularly, the SFR was estimated from the \iras\ \ensuremath{L_{IR}} and from \galex\ FUV, which provides an estimation of the unobscured and obscured SFR respectively\cite{Howell_2010}. The stellar mass was estimated from photometry using IRAC 3.6\ensuremath{\mu}m and Two Micron All Sky Survey (2MASS) K-band photometry.

For 21 galaxies, we retrieve their stellar masses and SFRs from the {\sc GALSPEC} Data Release 8\cite{Kauffmann_2003, Brinchmann_2004, Tremonti_2004}. Particularly, these quantities were retrieved following a Bayesian technique to match two stellar absorption indices, the \ensuremath{D_{n}(4000)} break\cite{Balogh_1999} and the H\ensuremath{\delta_{A}} Balmer absorption-line index\cite{Worthey_Ottaviani_1997}, from a library of different star formation histories from Monte Carlo realizations\cite{Kauffmann_2003}.

For ten galaxies, we retrieved their stellar masses\cite{Parkash_2018} derived from \wise\ bands W1 and W2 following the GAMA-derived stellar mass-to-light ratio relation\cite{Cluver_2014}. Their SFRs were obtained from \wise\ bands W1 and W3 to estimate the Balmer-decrement-corrected H\ensuremath{\alpha} luminosity\cite{Cluver_2017, Brown_2017, Parkash_2018}.

For seven galaxies, we retrieved their stellar masses from the S\ensuremath{^{4}}G\cite{Sheth_2010, Munoz-Mateos_2013, Querejeta_2015, Munoz-Mateos_2015} sample, which were estimated from the 2MASS photometry\cite{Bell_2003}. For these seven galaxies, we could not retrieve their SFRs. For another four galaxies, we obtained their stellar masses from the catalog published for a subsample of galaxies from the CASSIS database whose stellar masses and SFRs were obtained after applying the photometric SED fitting code {\sc CIGALE}\cite{Ciesla_2015, Boquien_2019} to their UV and near-IR observations\cite{Vika_2017}.

\subsection{Merger stage}
The GOALS sample also offers\cite{Stierwalt_2013} an estimation of the merger stage based on observations via IRAC 3.6\ensuremath{\mu}m images. By visual inspection, the authors provide a classification in the following categories: {\it nonmergers} when there is no sign of merger activity or its neighborhood lacks massive galaxies; {\it pre-mergers}, if the images reveal a pair of galaxies prior to the first encounter; {\it early-stage mergers}, pair of galaxies after the first encounter; {\it mid-stage mergers}, if they show amorphous disks, tidal tails and/or other signs of merger activity; and {\it late-stage mergers}, if they show a common envelope. 

\section{Inflow models}
\label{sec:models}
\subsection{Simulations of merging galaxies}
According to simulations of merging massive galaxies, a decrease in the metal content of galaxies, happening shortly after the first pericentre passage\cite{Rupke_2010}. When star-formation and metal enrichment from supernovae are included\cite{Montuori_2010}, the dilution is reported to be correlated to the star formation enhancement, and as the merger advances, the dilution is reduced and chemical enrichment becomes dominant. From this last model, the variations in metallicity as well as SFR are considered relatively to the initial conditions on the galaxies (see Figure 4\cite{Montuori_2010}). In this work, we considered the trends followed by galaxies in interaction with different initial conditions: Z\ensuremath{_{ini}}=0.8Z\ensuremath{_{\odot}}, Z\ensuremath{_{ini}}=0.4Z\ensuremath{_{\odot}} and Z\ensuremath{_{ini}}=0.2Z\ensuremath{_{\odot}}.

\subsection{Inflow model within galaxy evolution}
Another model that we consider to reproduce the drop of metallicity in some of the ULIRGs is based on basic equations of chemical evolution. For a standard galaxy, with original stellar mass M\ensuremath{_{*,0}}, SFR \ensuremath{\psi}, metallicity Z and with \ensuremath{\alpha } being the fraction in mass of long-lived stars and p the yield of metals (mainly oxygen), we can describe the evolution of gas mass ($G$), mass of long-lived stars ($S$) and mass of metals (\ensuremath{\mathcal{Z}}) following the equations\cite{Koppen_Edmunds_1999}:
\begin{equation}
\label{ec:G} \dot{G} = - \alpha \psi
\end{equation}
\begin{equation}
\label{ec:S} \dot{S} =  \alpha \psi
\end{equation}
\begin{equation}
\label{ec:Z} \dot{\mathcal{Z}} = - Z \alpha \psi + p \alpha \psi
\end{equation}
If a flow A of gas (in either direction) is introduced in the above equations, following the prescriptions\cite{Koppen_Edmunds_1999}, then Equation \ref{ec:G} is modified as:
\begin{equation}
\label{ec:G_new} \dot{G} = A - \alpha \psi
\end{equation}
Thus, we can express Equation \ref{ec:Z} in terms of the metal content (Z) as:
\begin{equation}
\label{ec:Z_new}  \ G\dot{Z} = p\alpha \psi - A Z
\end{equation}
For our model, we considered that both A and \ensuremath{\psi } present constant values for most of the evolving time of the galaxy, although to reproduce the massive inflow, we consider that they experience a peak (parameterized as a gaussian distribution), separated by 200 Myr (\ie the inflow A occurs before the peak in SFR), and we consider that the duration of the infall phase is bigger than the peak of star formation rate (which corresponds to 100 M\ensuremath{_{\odot }}yr\ensuremath{^{-1}}). After 1 Gyr of the process, we assume that the effects of outflows are significantly higher, changing the sign of A (from positive to negative). The value of \ensuremath{\alpha} is computed from the Initial Mass Function\cite{Kroupa_2001}, and considering long-lived stars those whose masses range from 0.01\ensuremath{_{\odot }} to 3\ensuremath{_{\odot }}. Finally, we assume a conservative yield p=0.003\cite{Pilyugin_2007}. As initial conditions, we assume a galaxy with stellar mass 10\ensuremath{^{11}} M\ensuremath{_{\odot}}, mass gas according to the observed ratio\cite{Parkash_2018}, and metallicity Z=0.84Z\ensuremath{_{\odot}} according to the obtained plateau in these study for massive galaxies.

Overall, according to this model, the process lasts for 2 Gyr: during which secular evolution is considered only in the first 200 Myr (i.e. both A and \ensuremath{\psi} show low and constant values). From 300 Myr to 900 Myr, the infall of gas dilutes the metallicity and increases the stellar mass. Finally, after 1 Gyr, the combination of stars polluting the ISM with metals and outflows contribute to enrich the metal content, while stellar mass shows little increase as the peak of star formation has already occurred. The infall process involves a total gas mass of 2.6\ensuremath{\cdot 10^{9}} M\ensuremath{_{\odot }}, which is equivalent to 3\% of the stellar mass of the galaxy and 15\% of its gas reservoirs.

\clearpage

\begin{center}
\bf{\large Methods References}
\label{M_Ref}
\end{center}

\section{Data Availability Statement}
Data supporting this study will be publicly available at the CDS.

\end{document}